\documentclass[journal]{IEEEtran}

\usepackage{amsmath,amsthm}
\usepackage{amssymb}
\usepackage{amscd}
\usepackage{textcomp}
\usepackage{fancyvrb}
\usepackage{wasysym}
   
\usepackage{graphicx}

\usepackage{listings}    

\usepackage{color}
\definecolor{red}{rgb}{1,0,0}
\definecolor{blue}{rgb}{0,0,1}
\definecolor{green}{rgb}{0,0.5,0}
\definecolor{magenta}{rgb}{1,0,1}

\newsavebox{\ieeealgbox}

\newcommand{\be}{\begin{equation}}
\newcommand{\ee}{\end{equation}}
\newcommand{\bea}{\begin{eqnarray}}
\newcommand{\eea}{\end{eqnarray}}
\newcommand{\bal}{\begin{align}}
\newcommand{\eal}{\end{align}}
\newcommand{\nn}{\nonumber}
\newcommand{\eye}{\mbox{$\mbox{1}\!\mbox{l}\;$}}

\renewcommand{\vec}[1]{\boldsymbol{#1}}

\begin{document}
\title{A Dual Method for Computing Power Transfer Distribution Factors }

\author{Henrik~Ronellenfitsch,%
\thanks{H. Ronellenfitsch is with the Max Planck Institute for Dynamics and Self-Organization (MPIDS), 37077 G\"ottingen, Germany, and
the Department of Physics and Astronomy, University of Pennsylvania, Philadelphia, PA 19104, USA}
Marc Timme,%
\thanks{M. Timme is with the Max Planck Institute for Dynamics and Self-Organization (MPIDS), 
 37077 G\"ottingen, Germany and the Faculty of Physics, University of G\"ottingen, 
37077 G\"ottingen, Germany}
Dirk~Witthaut%
\thanks{D. Witthaut is with the Forschungszentrum J\"ulich, Institute for Energy and Climate Research -
	Systems Analysis and Technology Evaluation (IEK-STE),  52428 J\"ulich, Germany
and the Institute for Theoretical Physics, University of Cologne, 
50937 K\"oln, Germany}}


\markboth{Journal of \LaTeX\ Class Files,~Vol.~13, No.~9, September~2014}%
{Shell \MakeLowercase{\textit{et al.}}: Bare Demo of IEEEtran.cls for Journals}

\maketitle

\begin{abstract}
    Power Transfer Distribution Factors (PTDFs) play a crucial role in
    power grid security analysis, planning, and redispatch.
    Fast calculation of the PTDFs is therefore of great importance.
    In this paper, we present a non-approximative dual method 
    of computing PTDFs.
    It uses power flows along topological cycles of the network
    but still relies on simple matrix algebra.
    At the core, our method 
    changes the size of the matrix that
    needs to be inverted to calculate the PTDFs from $N\times N$,
    where $N$ is the number of buses, to $(L-N+1)\times (L-N+1)$,
    where $L$ is the number of lines and $L-N+1$ is the number of
    independent cycles (closed loops) in the network while
    remaining mathematically fully equivalent.
    For power grids containing a relatively small number of 
    cycles,
    the method can offer a speedup of numerical calculations.
\end{abstract}


\IEEEpeerreviewmaketitle


\section{Introduction}
The supply of electric power is essential for the function of the economy
as well as for our daily life. Because of their enabling function 
for other infrastructures such as traffic or health care, power systems
are considered to be uniquely important
\cite{Amin05,Pour06,Krog08,Vleu10}.
The rise of renewable power sources puts new challenges to grid
operation and security, as they are typically strongly fluctuating
and often located far away from the load centers such that power
must be transported over large distances \cite{Amin05,Heid10,12powergrid,Pesc14}.
Thus efficient numerical methods are of great importance to 
analyze and improve the operation of power grids.

An important method to assess the security of a power grid and to
detect impending overloads is given by the linear sensitivity 
factors \cite{Wood14}.
Power transfer distribution factors (PTDFs) describe how the
real power flows change if power injection is shifted from one node
to another. Correspondingly, line outage distribution factors (LODFs) 
describe the flow changes when one line fails. These elementary
distribution factors can be generalized to reactive power flow \cite{Lee92} 
and multiple line outages \cite{Gule07}.
PTDFs and LODFs are heavily used in the planning, monitoring
and analysis of power systems \cite{Wood14}, for instance for
security analysis and contingency screening \cite{Saue81,Ng81,Stot87},
island detection \cite{Gule07b}, the simulation of cascading failures \cite{Dobs02}, 
transmission congestion management \cite{Sing98}, the estimation of available 
transfer capabilities \cite{Grav99} and re-dispatching in case of impending 
overloads \cite{Stot79,Mari05}.


The numerical simulation of large interconnected power systems can be 
computationally demanding. A particularly demanding step in the 
calculation of PTDFs is the inversion of the nodal susceptance matrix whose 
size is given by the number of buses $N$ in the grid. Computation times
can be crucial when many different load configurations of the 
grid have to be considered or for real-time security analysis.

In this paper we propose a new approach to computing PTDFs in
linear sensitivity analysis.
It uses a decomposition of power flows into cycle flows. Suppose we
inject some real power at node $s$ and take it out at node $r$.
We can satisfy real power balance by sending the power along an 
arbitrary path from $s$ to $r$. Obviously, this does not yield the physical
solution and must be corrected by flows over alternative paths from $s$ 
to $r$. Our analysis offers a systematic way to obtain the physically 
correct solution
by adding cycle flows which do not affect the power balance. 
This approach yields a novel method for calculating the PTDFs which
can be more efficient than established alternatives. In particular, the 
size of the matrix which has to be inverted is given by the number 
of independent cycles in the network, which is often much 
smaller than $N$.

\section{The DC approximation and linear sensitivity analysis}

The operation of power grids is determined by the conservation 
of real and reactive power, also called Tellegen's theorem \cite{Grai94,Wood14}.
The real power balance at one of the nodes $n = 1,\ldots,N$ reads
\begin{align}
   P_n = g_{nn} V_n^2 - \sum_{k \neq n} & \left( V_n V_k g_{nk} \cos(\theta_k - \theta_n) \right. \nn  \\
                                 & \left. + V_n V_k b_{nk} \sin(\theta_k - \theta_n) \right),
                                 \label{eq:loadflow-real}
\end{align}
where $P_n$ is the real power injection, i.e. the difference of generation 
and demand at the node $n$. The nodal voltage has magnitude $V_n$ and the
phase angle $\theta_n$. The nodes are connected by transmission lines or 
transformers with conductance $g_{nk}$ and susceptance $b_{nk}$.
Within the DC approximation one neglects ohmic losses, $g_{nk} = 0$, 
assumes that the voltage magnitude remains fixed and linearizes the sine 
function \cite{Grai94,Wood14}.
In the simplest case one expands the sine around the `empty' grid 
$\sin(\theta_k - \theta_n) \approx \theta_k - \theta_n$ and sets
all voltages magnitudes to 1 p.u. The nodal voltage
angles are then determined by the linear set of equations
\be
    \vec B \vec \theta  = \vec P
    \label{eq:dcapprox}
\ee
where $\vec B \in \mathbb{R}^{N \times N}$ is the nodal susceptance matrix
with elements
\begin{equation}
  B_{nk} = \left\{ 
   \begin{array}{lll}
   \displaystyle\sum \nolimits_{j=1}^{N} b_{n j} &  \mbox{if} & k = n; \\ [2mm]
     - b_{nk} & \mbox{if} & k \neq n.
   \end{array} \right.
\end{equation}
The vectors $\vec \theta = (\theta_1,\ldots,\theta_N)^t \in \mathbb{R}^N$ 
and $\vec P = (P_1,\ldots,P_N)^t \in \mathbb{R}^N$ summarize the nodal 
voltage angles and the real power injections, respectively. Here and in the 
following the superscript `$t$' denotes the transpose of a vector or matrix.
The real power flow from node $k$ to $n$ is then given by 
$F_{kn} = b_{kn} (\theta_k - \theta_n) = - B_{kn} (\theta_k - \theta_n) $.

We now consider an increase of the real power injection at node
$s$ and a corresponding decrease at node $r$ by the amount $\Delta P$.
The new vector of real power injections is given by 
\be
  \vec P' = \vec P + \Delta P\,\vec q_{sr},
  \label{eqn:real-power-balance}
\ee
where the components of $\vec q_{sr} \in \mathbb{R}^N$ 
are $+1$ at position $s$, $-1$ at position $r$ and zero otherwise.
In the interest of reducing notational clutter, we omit the explicit
dependence of $\vec P'$ on $s$ and $r$.
The nodal voltage angles then change by
\be
  \Delta \vec \theta = \Delta P \, \vec X \vec q_{sr},
  \label{eqn:delta-theta}
\ee
where $\vec X$
is the Moore-Penrose pseudo inverse of the nodal susceptance matrix,
\be
   \vec X = \vec B^{-1} \, .
\ee
Again, we omit the explicit dependence on $s$ and $r$.
It is noted that $\vec B$ is a  Laplacian matrix, which has one zero eigenvalue 
with eigenvector $(1,1,\ldots,1)^t$ \cite{Newm10}. This eigenvector 
corresponds to a global shift of the voltage angles which has no physical 
significance. Finally, the real power flows change by
$\Delta F_{ij} = b_{ij} (\Delta \theta_i - \Delta \theta_j)$ and
the associated power transfer distribution factors are given 
by \cite{Wood14},
\begin{align}
   \mbox{PTDF}_{(i,j),s,r} &:= \frac{\Delta F_{ij}}{\Delta P} \nn \\
    &  = b_ {ij} (X_{is} - X_{ir} - X_{js} + X_{jr}) .
    \label{eq:ptdf-definition}
\end{align}

Line outage distribution factors describe how the power flows change when
a line $(s,r)$ is lost. They are defined as \cite{Wood14}
\be
   \mbox{LODF}_{(ij),(sr)} = \frac{\Delta F_{ij}}{F_{sr}^{(0)}}
\ee
where the superscript $(0)$ denotes the flow before the outage.

The LODFs can be expressed by PTDFs in the following way. To consistently
model the outage of line $(s,r)$, one assumes that the line is disconnected 
from the grid by circuit breakers and that some fictitious real power 
$\Delta P$ is injected at node $s$ and taken out at node $r$. The entire flow 
over the line $(s,r)$ after the opening thus equals the fictitious injections 
$F'_{sr} = \Delta P$. 
Using PTDFs, we also know that
\begin{align}
   F'_{sr} = F^{(0)}_{sr}  + \mbox{PTDF}_{(s,r),s,r} \, \Delta P
\end{align}
Substituting $F'_{sr} = \Delta P$ and solving for $\Delta P$ yields
\begin{align}
 \Delta P = F'_{sr} = \frac{F^{(0)}_{sr}}{1 - \mbox{PTDF}_{(s,r),s,r}}
\end{align}
The change of real power flows of all other lines is given by
$\Delta F_{ij} = \mbox{PTDF}_{(i,j),s,r} \Delta P$ such that we
finally obtain 
\be 
    \mbox{LODF}_{(ij),(sr)} =\frac{\mbox{PTDF}_{(i,j),s,r}}{1 - \mbox{PTDF}_{(s,r),s,r}} \, .
\ee
 
The accuracy of the DC approximation and correspondingly the PTDFs and LODFs can 
be increased if one linearizes around a solved AC power flow base case, i.e. one
linearizes the equations (\ref{eq:loadflow-real}) around one particular solution.
This approach leads to so-called `hot-start DC models' or `incremental DC models'
\cite{Bald03,Bald05,Stot09}. The governing equation of these advanced DC models 
is still given by (\ref{eq:dcapprox}), but $\vec P$ and $\vec \theta$ now describe
the change of the power injections and phase angles with respect to the base case.
The matrix $\vec B$ explicitly depends on the base case, such that a new matrix has
to be inverted for every base case under consideration to compute the PTDFs and LODFs.
If many such base cases need to be analyzed, any speedup of the computation
can be extremely valuable.

\section{From edge space to cycle space}
In this section, we review some basic linear algebraic methods from
graph theory that we use in the rest of the paper.
We mainly follow \cite{Dies10} and \cite{Newm10}.

\begin{figure}[tb]
    \centering
\includegraphics[width=5cm]{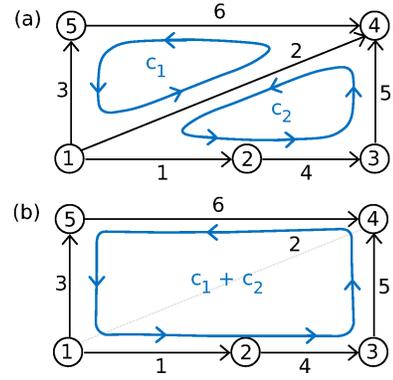}
\caption{Adding oriented cycles corresponds to the operation of symmetric
difference. (a) A basis of the cycle space $\mathcal C \simeq \mathbb R^2$ 
is given by the vectors $\vec c_1 = (0, 1, -1, 0, 0, 0, -1)^t$ and 
$\vec c_2 = (1, -1, 0, 1, 1, 0)^t$. Note that $L=6, N=5$ such that
$L-N+1 = 2$.
(b) A third cycle is obtained by forming the linear combination
$\vec c_3 = \vec c_1  + \vec c_2 = (1, 0, -1, 1, 1, -1)$.
Because the edges are oriented, this linear combination corresponds to
the symmetric difference of edge sets.
\label{fig:adding-cycles}}
\end{figure}

The connectivity structure of power transmission grids can be
modeled as a graph.
A graph $G=(V,E)$ consists of a set $V=\{v_1,\dots v_N\}$ of nodes 
(buses) and a set
$E=\{e_1,\dots e_L\}$ of edges (transmission lines or branches), 
where each $e_\ell \in E$ connects
two nodes, $e_\ell = \{ v_{\ell_1}, v_{\ell_2}\}, v_{\ell_{1,2}} \in V$.
Choosing an arbitrary but fixed orientation of the edges,
the graph can be encoded in terms of the node-edge incidence matrix 
$\vec I \in \mathbb{R}^{N \times L}$ \cite{Newm10} with components
\be
   I_{n,\ell} = \left\{
   \begin{array}{r l}
      1 & \; \mbox{if node $n$ is the tail of edge $e_\ell$},  \\
      - 1 & \; \mbox{if node $n$ is the head of edge $e_\ell$},  \\
      0     & \; \mbox{otherwise}.
  \end{array} \right.
\ee
The vector space $\mathcal V \simeq \mathbb R^N$ is called the 
\emph{node space}
of $G$ and the vector space $\mathcal E \simeq \mathbb R^L$ is called
the \emph{oriented edge space} of $G$.
$\mathcal V$ is spanned by the basis vectors $\vec v_i = 
(0,\dots, 0,1,0,\dots 0)^t \in \mathbb R^N$
with a $1$ at the $i$th position and zeros everywhere else.
Each basis vector $\vec v_i\in\mathcal V$ is associated with the node 
$v_i \in V$.
Similarly, the edge vector space $\mathcal E$ is spanned by the basis 
vectors
$\vec e_\ell = (0,\dots, 0,1,0,\dots 0)^t \in \mathbb R^L$, with one $1$ at
the $\ell$th position and is associated to the oriented edge $e_\ell \in E$.

The nullspace $\ker\vec I$ of the node-edge incidence matrix is called 
the \emph{cycle space} $\mathcal C$ and its elements consist
of all closed oriented paths (cycles) of $G$. Each cycle path
is represented by a vector $\vec c \in \mathbb R^L$ containing a $+1$
(if the edge oriented in the same way as the cycle)
or a $-1$ (if the edge is oriented in the opposite way from
the cycle) for each edge that is part of the cycle.
It can be shown that
$\mathcal C \simeq \mathbb R^{L-N+k} \subset \mathbb R^L$, 
where $k$ is the number of connected components of $G$ \cite{Dies10}. 
In the following, we will consider the case $k=1$ without loss of 
generality because each connected component can be analyzed separately.
Thus, there exists a basis of $L-N+1$ cycles from which
all other cycles can be obtained by (integer) linear combination.
Forming a linear combination with coefficients in $\{-1, 0, 1\}$ such that
edges contained in more than one cycle cancel is 
equivalent to the set-theoretic 
operation of symmetric difference \cite{Dies10} between the edge sets 
making up
the cycles (see Fig.~\ref{fig:adding-cycles}).
The symmetric difference of two sets $A$ and $B$ is defined as the set
$(A \cup B) \setminus (A \cap B)$. It contains all elements that are
contained in either $A$ or $B$ but not in both.

One cycle basis for $G$ is constructed as follows. Let $\mathcal T$ 
be a minimum
spanning tree of $G$. We note that such a minimum spanning tree contains
exactly $N-1$ edges.
For each edge $e\notin \mathcal T$, a cycle is defined by the set 
of $e$ together with the path in $\mathcal T$ connecting the nodes of $e$.
There are exactly $L-N+1$ such cycles, one for each edge not in $\mathcal T$.
They are linearly independent because each of them contains at least
one edge not found in the others ($e$), thus they make up a basis
of $\mathcal C$.

Given such a basis of cycles $\{c_1,\dots, c_{L-N+1}\}$,
we can now define the cycle-edge incidence matrix
$\vec C \in \mathbb{R}^{L \times (L-N+1)}$ by
\be 
   C_{\ell, c} = \left\{
   \begin{array}{r l}
      1 & \; \mbox{if edge $\ell$ is element of cycle $c$},  \\
      - 1 & \; \mbox{if reversed edge $\ell$ is element of cycle $c$},  \\
      0     & \; \mbox{otherwise}.
  \end{array} \right.
  \label{eqn:cycle-edge-matrix}
\ee
An explicit calculation shows that the matrix product
\be
\vec I \vec C = \vec 0 \in \mathbb R^{N\times (L-N+1)}.
\ee

\section{Dual method}
In this section, we introduce an alternative approach to network flows
based on the dual representation of the network. This method can speed up 
computations considerably and also sheds some light on topological aspects
of power flows. We note that the term duality here refers to the fact that
the PTDFs can be calculated equivalently using voltage angles or
cycle flows. However, in power grids with few cycles, there are much fewer
cycle flows to compute than voltage angles, thus making the cycle flow
method more economical.
To start, we reformulate the DC model in a
in a compact matrix notation.

Consider a grid with $N$ nodes
and $L$ transmission lines or transformers. The real power injections $\vec P$ 
and the voltage angles $\vec \theta$ are associated with the nodes of the network,
i.e. they are elements of $\mathbb{R}^N$. In contrast, power flows and 
susceptances are associated with lines, i.e they are represented by 
elements of $\mathbb{R}^L$. To be precise, 
we label all transmission lines by $\ell = 1, \ldots, L$. We then have a mapping 
between $\ell$ and an ordered pair of nodes $(i,j)$. The ordering is arbitrary but
must kept fixed as we are dealing with directed quantities such as power flows.
This mapping is encoded in the node-edge incidence matrix 
$\vec I \in \mathbb{R}^{N \times L}$.
Let $F_\ell$ denote the real power flows on the line $\ell$ and define 
the vector $\vec F = (F_1,\ldots,F_L) \in \mathbb{R}^L$. 
The susceptances $b_\ell$ of the transmission lines are summarized
in the branch susceptance matrix 
$\vec B_d = \mbox{diag} (b_1,\ldots, b_L) \in \mathbb{R}^{L\times L}$
and the branch reactance matrix
$\vec X_d = \mbox{diag} (1/b_1,\ldots, 1/b_L) \in \mathbb{R}^{L\times L}$
The nodal suceptance matrix then reads $\vec B = \vec I  \vec B_d \vec I^t$.

Within the DC approximation (\ref{eq:dcapprox}), the voltage angles and the flows 
are then written as
\begin{align}
   \vec \theta &= (\vec I  \vec B_d \vec I^t)^{-1} \vec P, \nn \\
   \vec F &= \vec B_d \, \vec I^t \vec \theta.
\end{align}
In practical applications it is common to define a PTDF matrix
which summarizes the distribution factors for all lines $\ell \in
\{1,\ldots,L\}$
and all nodes $r\in \{ 1\ldots,N\}$, fixing the slack node $s$. 
For notational
convenience we define the matrix $\vec S \in \mathbb{R}^{N\times N}$
generalizing the injection vectors $\vec q_{sr}$ used above,  
\be
   S_{ij} = \left\{
   \begin{array}{r l}
      - 1 & \; \mbox{if} \; i = j \neq s  \\
      + 1 & \; \mbox{if} \; i = s, j \neq s  \\
      0     & \; \mbox{otherwise}.
  \end{array} \right.
\ee
The PTDF matrix for a given slack node $s$ then reads    
\be
    \label{eq:ptdf-mat1}
   {\rm \bf PTDF} =  \vec B_d \vec I^t \, (\vec I \vec B_d \vec I^t)^{-1} \, \vec S \, .
\ee

For the calculation of the LODFs we do not fix a slack node $s$ but consider the case 
where power $\Delta P$ is injected at one end of a line $\ell$ and withdrawn at 
the other end. This is described by the matrix 
\be
   {\rm \bf PTDF}' =  \vec B_d \vec I^t \, (\vec I \vec B_d \vec I^t)^{-1} \, \vec I
\ee
such that the LODFs read
\be
   \label{eq:lodf-mat}
   {\rm \bf LODF}' =  {\rm \bf PTDF}' \, (\eye - {\rm diag}({\rm \bf PTDF}'))^{-1} \,  ,
\ee
where diag denotes the diagonal part of a matrix.

The standard approach to the calculation of PTDFs focuses on the nodes of the grid
and the computationally most demanding step is the inversion of the nodal susceptance
matrix $\vec B \in \mathbb{R}^{N\times N}$.
As an alternative, we propose a method that works with the real power flows
directly. Assume that additonal real power $\Delta P$ is injected at the slack node 
$s$ and taken out at node $r$. To find the change of the power flows we 
proceed in two steps. First, we construct all vectors $\Delta \vec F$ which satisfy 
the real power balance: The sum of all flows incident to a node must equal
the injected real power, $+ \Delta P$ at node $s$, $-\Delta P$ at node $r$ 
and zero otherwise, cf.~equation (\ref{eq:loadflow-real}).
In vectorial form this condition can be written as
\be
   \Delta P \vec q_{rs} \stackrel{!}{=} \vec I \vec \Delta \vec F 
\ee 
Then we single out the vector which yields the correct voltage 
angles $\Delta \vec \theta$, see \eqref{eqn:delta-theta}.

Any vector of flows $\Delta \vec F$ transporting the real power $\Delta P$ from 
node $s$ to node $r$ can be decomposed into two parts: a flow of magnitude
$\Delta P$ on an arbitrary path from node $s$ to node $r$ plus an arbitrary amount of 
cycle flows which do not affect the power balance at any node. This decomposition
is illustrated in Fig.~\ref{fig:5bus} (c) for a simple example network. 
    
The set of paths from a fixed slack node $s$ to  all other nodes in the grid
is referred to as a spanning tree in graph theory and can be calculated using
efficient algorithms \cite{Dies10}. A spanning tree is generally not unique; 
an arbitrary one can be chosen for our purposes. 
It is most convenient to encode the paths by a matrix 
$\vec T \in \mathbb{R}^{L\times N}$ with the components
\be 
  T_{\ell, r} = \left\{
   \begin{array}{r l}
      1 & \; \mbox{if line $\ell$ is element of path $s \rightarrow r$},  \\
      - 1 & \; \mbox{if reversed line $\ell$ is element of path $s \rightarrow r$},  \\
      0     & \; \mbox{otherwise}.
  \end{array} \right.
\ee
A power flow of magnitude $\Delta P$ from node $s$ to $r$  is then given by one vector 
$\Delta \vec F = \Delta P \vec T_{\cdot,r}$, where $\vec T_{\cdot,r}$ denotes the 
$r$th column of the matrix $\vec T$.

Furthermore, we need to characterize the cycle flows in the grid.
We denote the strength of the cycle flows by $f_c$ and define the vector 
$\vec f = (f_1\ldots,f_{L-N+1}) \in \mathbb{R}^{L-N+1}$, where $L-N+1$ 
is the number of independent cycles.
The flow vector $\Delta \vec F$ 
is then written as the direct flow and an arbitrary linear combination of cycle flows
as
\be
\Delta \vec F = \Delta P \, \vec T_{\cdot,r} + \vec C \vec f \, ,
   \label{eq:flow-picycle}
\ee
where $\vec C \in \mathbb R^{L-N+1}$ is the cycle-edge incidence matrix
\eqref{eqn:cycle-edge-matrix}.
For any choice of $\vec f$, $\Delta \vec F$ satisfies the real 
power balance at each node because $\vec I \vec C = \vec 0$.

\begin{figure}
    \centering
    \includegraphics[width=.4\textwidth]{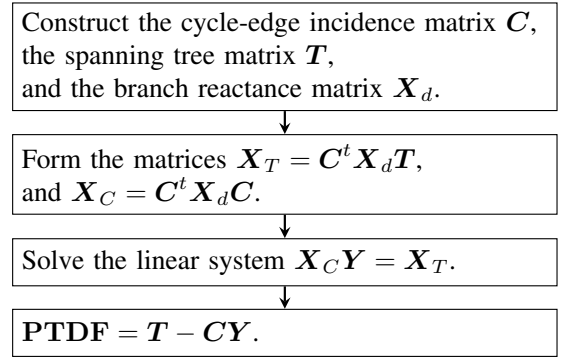}
    \caption{Flow chart describing how to use the dual method to
        compute the PTDF matrix from equation \eqref{eq:ptdf-mat2}.
    \label{fig:flow-chart}
    }
\end{figure}

In a second step, we determine the correct physical flow vector $\Delta  \vec F$.
This amounts to calculating the cycle flow strengths $\vec f$ such that all
voltage angles in the grid are unique. A necessary and sufficient condition
is that the sum of all angle differences along any closed cycle equals zero, 
\be
     \label{eq:phasecon}
     \sum_{(ij) \in {\rm cycle} \,  c}  \left( \Delta \theta_i - \Delta\theta_j\right) \stackrel{!}{=} 0 \, .
\ee
As the cycles form a vector space it is sufficient to check this condition for the $L-N+1$ basis
cycles. Using $\Delta F_{ij} = b_{ij} (\Delta \theta_i - \Delta\theta_j)$, the condition
reads
\be
    \sum_{(ij) \in {\rm cycle} \,  c} \Delta F_{ij}/b_{ij} \stackrel{!}{=} 0
\ee
for all basis cycles $c\in\{1,\dots,L-N+1\}$.
This set of equations can be recast into matrix form,
\be
   \vec C^t  \vec X_d \Delta \vec F = \vec 0.
\ee
Inserting the decomposition (\ref{eq:flow-picycle}) we obtain
\be
   \vec C^t \vec X_d \vec C \vec f + \Delta P \, \vec C^t  \vec X_d \vec T_{\cdot,r} = \vec 0, 
\ee
which can be solved for the cycle flows $\vec f$. The formal solution is
\be
    \vec f = - \Delta P  ( \vec C^t \vec B_d^{-1} \vec C )^{-1} \vec C^t  \vec B_d^{-1}  
          \vec T_{\cdot,r} \, .
          \label{eq:cycle-flow-solution}
\ee
The changes of the real power flows are given by equation (\ref{eq:flow-picycle}).
The PTDF matrix summarizing the distribution factors for all nodes $r$ and a
fixed slack bus $s$ is then calculated by inserting equation 
(\ref{eq:cycle-flow-solution}) into equation (\ref{eq:flow-picycle}) and
subsequently using the definition (\ref{eq:ptdf-definition}). The
result is
\be
    {\rm \bf PTDF} = 
    \left[ \eye - \vec C  ( \vec C^t \vec X_d \vec C )^{-1} \vec C^t  \vec X_d \right]  \vec T \, .
    \label{eq:ptdf-mat2}
\ee
An efficient way of using this formula is shown in the flow chart
Fig.~\ref{fig:flow-chart}.

For the calculation of the LODFs we do not need to calculate the matrix
$\vec T$, as only injections at the terminal end of the lines $\ell = 1,\ldots,L$
are considered. The LODF matrix is given by equation (\ref{eq:lodf-mat})
with
\be
    {\rm \bf PTDF}' = 
    \left[ \eye - \vec C  ( \vec C^t \vec X_d \vec C )^{-1} \vec C^t  \vec X_d \right] \, ,
    \label{eq:ptdf-mat2-diag}
\ee
where the derivation proceeds analogously to that of equation
(\ref{eq:ptdf-mat2}).

We stress that equation \eqref{eq:ptdf-mat2} is not an approximation to
the conventional equation \eqref{eq:ptdf-mat1}, it is an alternative
but mathematically fully equivalent way of computing the PTDFs.
There are no approximations involved in the transformation to the dual description. 
The applicability of linear distribution factors itself is discussed in 
\cite{Bald03,Bald05,Hert06,Stot09}.


\section{Example}

\begin{figure}[tb]
\includegraphics[trim=3.5cm 1.5cm 7cm 1cm, clip, width=7.5cm]{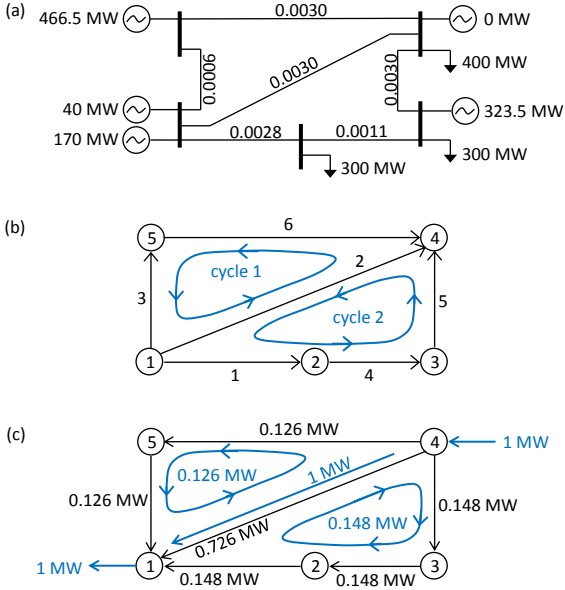}
\caption{
\label{fig:5bus}
Physical and cycle flows in a 5-bus example network
(a) Circuit diagram of the 5-bus example network \cite{matpower}. The
reactances of each line are given in p.u.
(b) Topology of the network. Labels of nodes, lines and cycles as used
in the text.
(c) Calculation of PTDFs. Black numbers give the physical power flow
$\Delta \vec F$ when 1 MW is injected at node 4 and withdrawn at node 1.
The flows $\Delta F$ can be decomposed into a 1 MW-flow from node 4 to 
node 1 on the direct path plus two cycle flows shown by blue arrows.
}
\end{figure}

As an instructive example we consider the 5-bus test grid from MATPOWER \cite{matpower}
with $N=5$ and $L=6$. The circuit diagram as well as the topology of the grid are 
illustrated in Fig. \ref{fig:5bus}. The node-edge incidence matrix is given by 
\be
    \vec I = \begin{pmatrix}
       +1 & +1 & +1 & 0 & 0 & 0 \\
       -1 & 0 & 0 & +1 & 0 & 0\\
       0 & 0 & 0 & -1 & +1 & 0 \\
       0 & -1 & 0 & 0 & -1 & -1 \\
       0 & 0 & -1 & 0 & 0 & +1 \\
       \end{pmatrix}.
\ee
The grid contains $2$ independent cycles, which are chosen as\\
\phantom{000} cycle 1: line 2, reverse line 6, reverse line 3. \\
\phantom{000} cycle 2: line 1, line 4, line 5, reverse line 2\\
The cycle-edge incidence matrix thus reads
\be
  \vec C^t = \begin{pmatrix}
        0 & +1 & -1 & 0 & 0 & -1\\
       +1 & -1 & 0 & +1 & +1 & 0 \\
       \end{pmatrix}.
\ee

Fig. \ref{fig:5bus} (c) shows the physical real power flows as well as the 
cycle decomposition (\ref{eq:flow-picycle}) for $s=4$ and $r=1$.
There is a direct flow of magnitude $\Delta P = 1 \, {\rm MW}$ from node 
4 to node 1. Additional cycle flows along the two independent cycles do
not affect the power balance. The physical flows are recovered
when the strength of the cycle flows is given by
$f_1 = 0.126$ MW and $f_2 = -0.148$ MW.

The example shows how the new method can potentially speed-up
the computation of PTDFs. The conventional approach focuses on the
$N=5$ nodes of the grid and calculates how the voltage angles change.
But these $N=5$ variables are not independent but related topologically
through the condition (\ref{eq:phasecon}).
In the new approach only $2$ independent variables, the cycle flow
strengths, must be calculated. The changes in power flow then
follow directly from \eqref{eq:flow-picycle} and the PTDFs
from \eqref{eq:ptdf-mat2}.

\section{Implementation and Computational Performance}
The computationally most demanding part in the calculation of 
PTDFs is the inversion of a large matrix. In the conventional approach
defined by equation (\ref{eq:ptdf-mat1}), the $N \times N$-matrix
$\vec B = \vec I \vec B_d \vec I^t$ has to be inverted. 
The dual method defined by equation (\ref{eq:ptdf-mat2})
requires the inversion of a $(L-N+1) \times (L-N+1)$-matrix 
$\vec C^t \vec X_d \vec C$ instead.
In real-world power grids, the number of cycles 
$L-N+1$ is often much smaller than $N$. Hence a
much smaller matrix has to be inverted which can lead to
a significant speed-up of numerical calculations.

In practical applications, the formula (\ref{eq:ptdf-mat1}) for the conventional computation of PTDFs
is usually slightly modified. As noted before, the nodal susceptance matrix $\vec B$ has one
zero eigenvalue associated with a global shift of the voltage angles.
One generally fixes the voltage angle at the slack node $s$ at a value 
of zero and excludes this node from the calculation. Equation
(\ref{eq:ptdf-mat1}) then reads
\be
    \label{eq:ptdf-mat3}
   {\rm \bf PTDF}_{\rm red} =  \vec B_{f, \rm red} \, \vec B_{\rm red}^{-1},
\ee
where $\vec B_{f} = \vec B_d \vec I^t$ and the subscript `red'
indicates that slack bus is excluded, i.e. the $s$th row and column 
is deleted in $\vec B$ and the $s$th column for all other matrices.
Furthermore, one does not have to compute the full inverse of the
matrices but can solve a system of linear equations instead. 
For instance, one can solve 
\be
   {\rm \bf PTDF}_{\rm red} \vec B_{\rm red} =  \vec B_{f, \rm red} \, .
\ee
for $\vec B_{\rm red}$ instead of computing the inverse in equation 
(\ref{eq:ptdf-mat3}). This approach is implemented for instance in the
popular software package MATPOWER 5.1 \cite{matpower}.

\begin{figure}[tb]
\begin{Verbatim}[frame=single]
tic;
PTDF1 = zeros(L,N);
Bf = Bd * I';
Bbus = I * Bf;
PTDF1(:,an) = full(Bf(:,an)/Bbus(an,an));
toc
     
tic;
Xf = C' * Xd;
Xc = Xf * C;
Xt = Xf * T;
PTDF2 =  T - C * full(Xc \ Xt);
toc
\end{Verbatim}
\caption{\textsc{Matlab} code to compare the runtime of the conventional
algorithm and the dual method. All variable names are same as those
used in the text with the exception of \texttt{an}, which indexes all
nodes except for the slack.}
\label{fig:algo}
\end{figure}

The dual method yields the formula (\ref{eq:ptdf-mat2}) for the computation of the PTDF matrix. Again one can omit the full matrix inversion and solve a linear system of equations instead. Then computation is then done in two steps
\bea
    {\rm Solve} \; && (\vec C^t \vec X_d \vec C) \, \textbf{TEMP} = (\vec C^t \vec X_d \vec T) \nn \\
     {\rm Compute} \; && \textbf{PTDF} = \vec T - \vec C \; \textbf{TEMP} .
     \label{eq:ptdf-mat2-noinv}
\eea

If one is only interested in calculating LODFs by means of equations
(\ref{eq:lodf-mat}) and (\ref{eq:ptdf-mat2-diag}), 
a further speedup is possible by defining 
$\vec {\tilde C}^t = \vec C^t \sqrt{\vec X_d} = \vec Q \vec R$ using a 
$QR$ decomposition. Then,
\be
{\rm \bf PTDF}' =     
\left[ \eye - \sqrt{\vec B_d} \vec Q \vec Q^t \sqrt{\vec X_d} \right],
\ee
completely eliminating the need for inverting any matrices.

\subsection{Sparse numerics}

\begin{table*}[t!]
\centering
\caption{
\label{tab:sparse}
Comparison of CPU time for the calculation of the PTDFs
obtained with \textsc{Matlab} sparse matrices.
}
\begin{tabular}{ |c|c|c|c|c|c|c|c|c| }
  \hline
  \multicolumn{2}{|c|}{Test Grid}  
         & \multicolumn{4}{|c|}{Grid Size}
         & \multicolumn{2}{|c|}{CPU time in seconds} 
         & speedup \\
  name & source &
      nodes & lines & cycles & cycles/nodes &
     Conventional method & Dual method & \\
  & & $N$ & $L$ & $L-N+1$ & $\frac{L-N+1}{N}$ &   
       Eq.~(\ref{eq:ptdf-mat3}) &  Eq.~(\ref{eq:ptdf-mat2-noinv}) & 
   $t_{(32)}/t_{(34)}$   \\  
  \hline
   Transmission grids: & & & & & & & & \\
 case300 & \cite{matpower} & 300 & 409 & 110 & 0.37 & $ 0.0038 \pm 0.0006 $ & $ 0.0020 \pm 0.0005 $ & $ 1.90 $ \\
 case1354pegase & \cite{Flis13} & 1354 & 1710 & 357 & 0.26 & $ 0.131 \pm 0.006 $ & $ 0.038 \pm 0.001 $ & $ 3.46 $ \\
 GBnetwork & \cite{GBnet} & 2224 & 2804 & 581 & 0.26 & $ 0.38 \pm 0.00 $ & $ 0.09 \pm 0.00 $ & $ 4.43 $ \\
 case2383wp & \cite{matpower} & 2383 & 2886 & 504 & 0.21 & $ 0.45 \pm 0.01 $ & $ 0.12 \pm 0.00 $ & $ 3.72 $ \\
 case2736sp &\cite{matpower} & 2736 & 3495 & 760 & 0.28 & $ 0.63 \pm 0.02 $ & $ 0.30 \pm 0.02 $ & $ 2.06 $ \\
 case2746wp & \cite{matpower} & 2746 & 3505 & 760 & 0.28 & $ 0.646 \pm 0.031 $ & $ 0.307 \pm 0.031 $ & $ 2.11 $ \\
 case2869pegase & \cite{Flis13} & 2869 & 3968 & 1100 & 0.38 & $ 0.709 \pm 0.052 $ & $ 0.224 \pm 0.002 $ & $ 3.16 $ \\
 case3012wp & \cite{matpower} & 3012 & 3566 & 555 & 0.18 & $ 0.696 \pm 0.052 $ & $ 0.173 \pm 0.014 $ & $ 4.04 $ \\
 case3120sp & \cite{matpower} & 3120 & 3684 & 565 & 0.18 & $ 0.735 \pm 0.045 $ & $ 0.184 \pm 0.020 $ & $ 3.99 $ \\
 westernus & \cite{Watt98} & 4941 & 6594 & 1654 & 0.33 & $ 1.906 \pm 0.079 $ & $ 0.669 \pm 0.054 $ & $ 2.85 $ \\
 case9241pegase & \cite{Flis13} & 9241 & 14207 & 4967 & 0.54 & $ 9.49 \pm 0.97 $ & $ 7.61 \pm 0.46 $ & $ 1.25 $ \\
 Distribution grids: & & & & & & & & \\
 bus\_873\_7 & \cite{reds} & 880 & 900 & 21 & 0.02 & $ 0.043 \pm 0.000 $ & $ 0.007 \pm 0.001 $ & $ 6.43 $ \\
 bus\_10476\_84 & \cite{reds} & 8489 & 8673 & 185 & 0.02 & $ 4.49 \pm 0.43 $ & $ 0.68 \pm 0.21 $ & $ 6.63 $ \\
\hline
\end{tabular}

\end{table*}

\begin{figure}[tb]
\includegraphics[width=8cm]{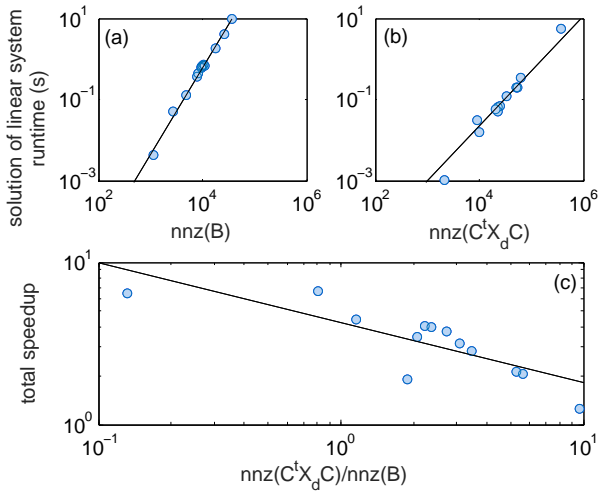}
\centering
\caption{\label{fig:speedup-s}
Depending on network topology, the dual method can significantly speed up the calculation of the
PTDFs. 
(a, b) The runtime of the linear inversion step using the \textsc{Matlab} Cholesky decomposition solver scales with the number of non-zero elements (`nnz') of the matrices $\vec B$ and the matrix $\vec C^t \vec X_d \vec C$, respectively. However, the scaling exponents and the prefactors are vastly different, such that the dual method is faster. 
(c) The total speed-up is given by the ratio of the runtimes of the conventional node-based method with fixed slack and the dual method including all matrix multiplications. The total speed-up lies between 1.25 and 6.63 for the test grids under consideration. The runtimes have been evaluated using the code listed in Fig.~\ref{fig:algo} and are listed in table \ref{tab:sparse}. The black lines are power-law fits to the data.
}
\end{figure}

We test how the dual method presented in this paper can speed up actual computations 
using several test cases. We compare the conventional method
using Eq.~(\ref{eq:ptdf-mat3}) to the dual method given by Eq.~(\ref{eq:ptdf-mat2-noinv}).
The runtimes of all methods are evaluated using the
\textsc{MATLAB} script listed in Fig.~\ref{fig:algo}. 
In addition, we evaluate the runtime 
for the solution of the linear set of equations alone, i.e. execution
of the commands
\texttt{Bf(:,an)/Bbus(an,an)} and
\texttt{Xc$\backslash{}$Xt}, respectively.
The variable \texttt{an} is a vector indexing all nodes except 
for the slack node. All other variables are the same as before.

Because all input matrices involved exhibit a sparse
structure (i.e., they contain many identically zero entries), it is
sensible to test computational performance using specialized
sparse numerics. To this end, we converted all input matrices appearing
in the code of Fig. \ref{fig:algo} into the sparse format used
by \textsc{Matlab} using the \texttt{sparse} command. 
Internally, \textsc{Matlab} then employs the high-performance
supernodal sparse Cholesky decomposition solver \textsc{Cholmod} 1.7.0
for the solution of the linear system of equations.
The resulting PTDF matrix is full (i.e. it usually contains no zeros), such that we converted
the the results back to a full matrix using the command \texttt{full}.
For the dual method, care has to be taken about where to do the 
conversion (see the code example in Fig. \ref{fig:algo}).

The results are shown in Table~\ref{tab:sparse} 
and Fig.~\ref{fig:speedup-s} for various test grids 
from \cite{matpower}, \cite{Flis13}, \cite{GBnet}, \cite{Watt98}  and \cite{reds}. 
For the sake of simplicity we have merged all parallel transmission lines, 
such that the graph contains no loops. Tests were carried out on a workstation 
with an Intel Xeon E5-2637v2 processor  at 3.5 Ghz and 256 GB RAM using 
Windows 8.1Pro, MATLAB version R2015a and MATPOWER version 5.1.
All results were averaged over 100 runs
and the standard deviation is given.

We find that the dual method (\ref{eq:ptdf-mat2-noinv}) significantly speeds up the computation
for all test grids under considerations. The dual method is faster by a factor of
up to 4.43 for transmission grids and up to 6.63 for distribution grids.
The speed-up is even more pronounced if we consider the solution of the linear system
only, ranging up to 12.08 for transmission grids and up to 75.03 
for distribution grids.
However, the dual methods requires additional matrix multiplications to construct the PTDFs,
which reduces the total speed-up.

\subsection{Dense numerics}
\begin{figure}[tb]
\includegraphics[width=8cm]{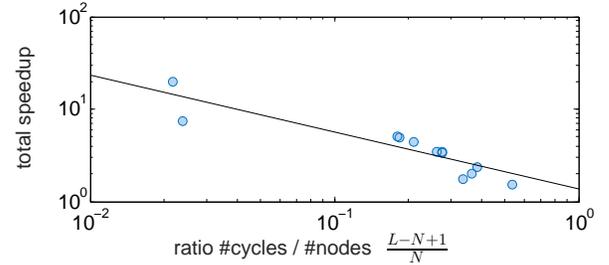}
\centering
\caption{\label{fig:speedup}
Total speed-up of the dual method using dense numerics using the same test grids as in
Fig.~\ref{fig:speedup-s} as a function of the ratio of the numbers of
cycles $L-N+1$ and the number of nodes $N$. 
The black line is a power law fit $\alpha \times [(L-N+1)/N]^{-\gamma}$ 
to the data, which yields the parameters 
$\alpha = 1.355$ and $\gamma = 0.616$.
}
\end{figure}

Traditional dense numerical performance is dominated by the 
dimensionality of the problem, which is given
by the number of nodes $N$ for the conventional method and the number
of cycles $L-N+1$ for the dual method. Hence, the ratio of the runtimes
(the speedup) is essentially determined by the ratio $(L-N+1)/N$. 
The speed-up obtained using the dual method is even larger than
in the case of sparse numerics and reaches up to a factor of 5.06
for the transmission test grids and 19.89 for the distribution test grids
studied here.
Numerically, we find that the total speedup scales as a power law 
with the ratio of the number of cycles and the number of nodes 
$(L-N+1)/N$ with an exponent $\gamma = 0.616$ 
(Fig.~\ref{fig:speedup}).

\section{Potential applications}

\subsection{Speeding up calculations}

The dual method can significantly speed up the calculation of PTDFs depending on the 
network topology as shown in the previous section. Thus it can be useful for time-critical 
applications where PTDFs must be calculated repeatedly -- for instance in `hot-start DC models' or
`incremental DC models'. In these cases the matrix $\vec B$ is different for all base cases 
under consideration and the distribution factors have to be calculated separately for
all cases (see \cite{Stot09} and references therein).
For instance, in the flow-based approach to capacity allocation and congestion management
in Europe, PTDFs must be calculated for each timeframe  \cite{Eu15}.

There are two main reasons making the dual method particularly suitable for this
type of applications:
First, the speedup can be significant but the absolute computation times are such that
the conventional method is also feasible when the computation time is not critical.
Second, the application of the dual method makes use of the spanning
tree $\vec T$ and the cycle incidence matrix $\vec C$, the calculation of
which also requires some computational resources. However, both $\vec C$ and
$\vec T$ depend only on the network topology, but not on the actual values of
$B_{nk}$. They are identical for all base cases such that they can be calculated once 
during initialization and stored for further use. Even more, they can be updated 
easily when a new bus or a transmission line is connected to the 
grid~\cite{Chin1978}.

The speed-up is even more pronounced for distribution grids, which are ultra-sparse by
construction. The use of PTDFs is less common in distribution grids, but has recently gained some
interest in the control of grid congestion due to electric vehicle charging \cite{OCon12,Li14}.

\subsection{Changes of the grid topology}

In addition to purely numerical benefits, the dual formulation can be used to derive analytical results on how power flows in complex grid topologies. For example, it shows in an intuitive way how the flows are affected by changes of the grid topology. To demonstrate this we consider the closing of a tie-switch in a distribution grid. Assuming that the grid was tree-like in its original configuration, the closing induces a single unique cycle $c$ and the cycle incidence matrix $\vec C$ reduces to a vector in $\mathbb{R}^L$. The PTDFs change as
\be
  \Delta \textbf{PTDF} = - \vec C  (\vec C^T \vec X_d \vec C)^{-1}  \vec C^T \vec X_d \vec T ,
\ee
which allows for a very simple interpretation. Consider the $r$th row of the PTDF matrix and assume that the root of the tree has been chosen as slack. Then we have
\be
   \Delta \textbf{PTDF}_{\cdot,r} = - \vec C \frac{\sum_{\ell \in {\rm cycle} \, c \, {\rm and} \, \ell \in \,
     {\rm path} \, s \rightarrow r} x_\ell}{\sum_{\ell \in {\rm cycle} \,  c} x_\ell}
\ee
This formulation shows two main aspects of flow rerouting due to the closing of the switch: First the PTDF matrix
changes only for the lines which are part of the induced cycle. Second, the strength of the change is given by the ratio of two sums of line reactances: In the denominator we sum over all lines which are part of the cycle $c$ and in the numerator we sum only over which are part of the cycle $c$ \emph{and} the direct path from the slack node $s$ to node $r$. Loosely speaking, this ratio measures the overlap of the induced cycle and the direct path from $s$ to $r$. This and similar results can also be obtained in a different way, but are immediately obvious in the dual formulation. 

\subsection{Quantifying unscheduled flows}

Unscheduled power flows or loop flows refer to the fact that power can flow through several paths in a meshed
grid, and thus lead to different flows than scheduled during trading. These flows significantly contribute to limits for limit cross border trading, e.g. in the interconnected European grid, and have played an important role in events like the 2003 North American blackout \cite{Bowe02,Sury08,Mari10}. We here discuss the quantification of unscheduled flows on a nodal level. Unscheduled flows between different loop flows zones can be treated in the same manor using zonal PTDFs and effective line parameters \cite{Purc05,Tran15}.

Suppose a generator at node $s$ sells power $P$ to a consumer at node $r$ and let $\vec \pi \in \mathbb{R}^L$ be a vector which encodes the scheduled path for the power flow. In the simplest case there is a direct connection between nodes $s$ and $r$ via the transmission line $\ell$. Then $\vec \pi$ is a unit vector which is one at position $\ell$ and zero otherwise. Formula (\ref{eq:ptdf-mat2}) now directly yields the actual, scheduled and unscheduled flows induced by this transaction:
\be
   \vec F_{\rm actual} = \underbrace{P \vec \pi}_{=: \vec F_{\rm scheduled}} 
   \underbrace{  - P \vec C  (\vec C^T \vec X_d \vec C)^{-1}  \vec C^T \vec X_d \vec \pi}_{=: \vec F_{\rm unscheduled}}.
\ee

\section{Conclusions}

Power Transfer Distribution Factors and Line Outage Distribution Factors
are important tools enabling the efficient planning of power grid
operations in contingency cases.
Especially for very large networks, computational efficiency can be crucial.
We presented a novel method of calculating the PDTFs based on cycle flows,
eliminating redundant degrees of freedom present in the conventional approach. 
The main step of the computation is the solution of a large linear system of equations,
whose dimensionality is reduced from the number of nodes $N$ to the number of
fundamental cycles $L-N+1$, $L$ being the number of branches.
This can result in a significant improvement
of the computation time 
for grids where the number of cycles $L-N+1$ is significantly 
smaller than the number of nodes $N$.
In addition, the cycle flow description provides
a conceptual advantage, dealing with power flows directly
without recourse to voltage angles.
We finally note that mathematically equivalent models of flow are used 
to describe hydraulic networks \cite{Hwan96} or vascular networks of plants 
\cite{Kati10}.

\section*{Acknowledgments}
We gratefully acknowledge support from 
the Helmholtz Association (via the joint initiative
``Energy System 2050 -- A Contribution of the Research Field Energy''
and the grant no.~VH-NG-1025 to D.W.) and
the Federal Ministry of Education and 
Research (BMBF grant nos.~03SF0472B and ~03SF0472E to D.W. and M.T.).
The work of H. R. was supported in part by the IMPRS Physics of Biological
and Complex Systems, G\"ottingen.

\bibliographystyle{IEEEtran}
\bibliography{ptdf}

%

\begin{IEEEbiographynophoto}{Henrik Ronellenfitsch}
    received his BSc and MSc in physics from ETH Z\"urich, Switzerland
    in 2010 and 2012, respectively. He completed his PhD studies
    at the Max Planck Institute for Dynamics and Self-Organization
    in G\"ottingen, Germany in 2016 working on problems involving 
    flow and transport networks as applied to biological and human-made
    systems and now works as a Postdoctoral researcher at the
    Department of Physics and Astronomy, University of Pennsylvania.
\end{IEEEbiographynophoto}

\begin{IEEEbiographynophoto}{Marc Timme}
    studied physics and applied mathematics at the Universities of W\"urzburg (Germany), Stony Brook (New York, USA) and G\"ottingen (Germany). He holds an M.A. in Physics from Stony Brook and a Doctorate in Theoretical Physics (G\"ottingen). After postdoctoral and visiting stays at Cornell University (New York, USA) and the National Research Center of Italy (Sesto Fiorentino) he is heading the Max Planck Research Group on Network Dynamics at the Max Planck Institute for Dynamics and Self-Organization. He is Adjunct Professor at the University of G\"ottingen. 
\end{IEEEbiographynophoto}

\begin{IEEEbiographynophoto}{Dirk Witthaut}
    received his Diploma (MSc) and PhD in Physics from the Technical University of 
    Kaiserslautern, Germany, in 2004 and 2007, respectively. He has been a Postodoctoral 
    Researcher at the Niels Bohr Institute in Copenhagen, Denmark and at the Max Planck 
    Institute for Dynamics and Self-Organization in G\"ottingen, Germany and a
    Guest Lecturer at the Kigali Institue for Science and Technology in Rwanda.
    Since 2014 he is heading a Research Group at Forschungszentrum J\"ulich, Germany.    
\end{IEEEbiographynophoto}

\end{document}